\documentclass{article}

\usepackage[english]{babel}
\usepackage{authblk}

\usepackage[letterpaper,top=2cm,bottom=2cm,left=3cm,right=3cm,marginparwidth=1.75cm]{geometry}

\usepackage{amsmath}
\usepackage{graphicx}
\usepackage[colorlinks=true, allcolors=blue]{hyperref}

\title{Full-aperture extended-depth oblique plane microscopy through dynamic remote focusing}

\author[1]{Paolo Pozzi\thanks{paolo.pozzi@polimi.it}}
\author[1]{Vipin Balan}
\author[1]{Alessia Candeo}
\author[2]{Alessia Brix}
\author[2]{Anna Silvia Pistocchi}
\author[1]{Cosimo D'Andrea}
\author[1]{Gianluca Valentini}
\author[1]{Andrea Bassi}
\affil[1]{Dipartimento di Fisica, Politecnico di Milano, Piazza Leonardo da Vinci 32, I-20133 Milano, Italy}
\affil[2]{Dipartimento di Biotecnologie Mediche e Medicina Traslazionale, Università degli Studi di Milano, Via Festa del Perdono 7 - 20122 Milano, Italy}

\begin{document}
\maketitle

\begin{abstract}
Oblique plane microscopy is a method enabling light-sheet fluorescence imaging through a single microscope objective lens by focusing on a tilted plane within the sample. To focus the fluorescence emitted by the oblique plane on a camera, the light is imaged through a pair of remote objective lenses, facing each other at an angle. The aperture mismatch resulting from this configuration limits the effective numerical aperture of the system, reducing image resolution and signal intensity.

This manuscript introduces an alternative method to capture the oblique plane on the camera. Instead of relying on angled objective lenses, an electrically tunable lens is employed. This lens adjusts the focal plane of the microscope synchronously with the rolling shutter of a scientific CMOS camera. In this configuration the entire aperture of the objective is effectively employed, increasing the resolution of the system. Moreover, a variety of objective lenses can be employed, enabling the acquisition of wider axial fields of view compared to conventional oblique plane microscopy.
\end{abstract}

\section{Introduction}
Light-sheet microscopy is the gold standard method for fluorescence imaging of complex three dimensional samples at high rates \cite{girkin2018_lightsheet_review}. Due to the need for multiple perpendicular microscope objectives around the sample, standard light-sheet microscopy has significant geometric constraints for the shape and mounting method of the observed sample. Although this three-dimensional arrangement of sample mounting can sometimes be an advantage\cite{candeo2017_lightsheet_arabidopsis, candeo2016_mouse_intestine}, its use with standard slides and Petri dishes, or in intravital applications in small rodents can be challenging, requiring complex geometries \cite{glaser2022_open_top_lightsheet} for the objective lenses and for sample positioning.

Oblique plane microscopy (OPM) \cite{dunsby2008_original_opm} is a variant of light-sheet microscopy, in which a single objective lens is used both for light-sheet illumination and fluorescence detection, greatly simplifying imaging of samples optically accessible from a single direction. To achieve this, the excitation light is confined to one edge of the objective aperture, so that the light sheet propagates at an angle with respect to the optical axis.
Since the illuminated plane is tilted from the optical axis, only a small portion of it would be in focus on a camera in a standard detection path. To have the oblique plane in focus within the whole field of view of the microscope, a re-projection setup is employed in the detection path, with a secondary microscope objective forming a low magnification image of the sample, and a tertiary objective observing such image at an angle.

\begin{figure}[ht!]
\centering\includegraphics[width=0.8\textwidth]{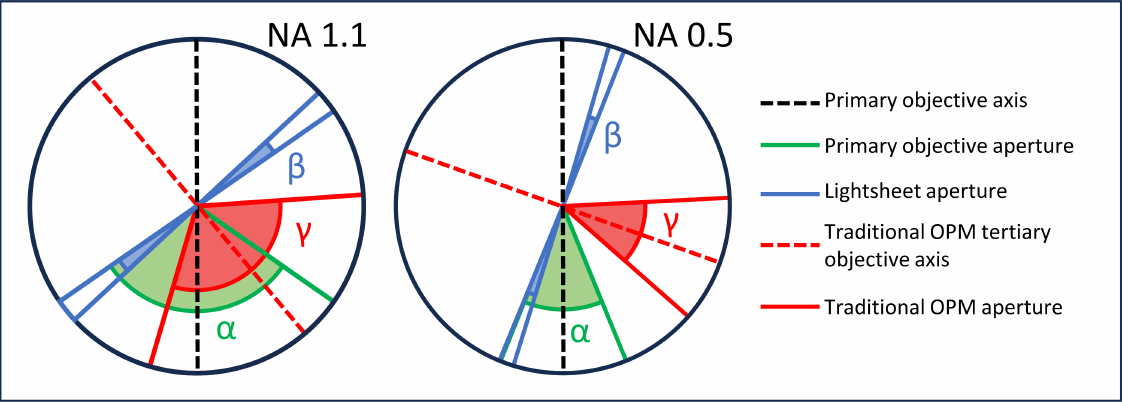}
\caption{\textbf{OPM apertures.} Representation of OPM apertures for high and low numerical aperture water immersion objectives. The angle $\alpha$ represents the aperture of the primary objective, $\beta$ is the light-sheet aperture, $\gamma$ is the aperture of an ideal tertiary objective in standard OPM with the same numerical aperture as the primary. The effective aperture of the system is the angle of the overlap between $\alpha$ and $\gamma$}\label{fig. Apertures}
\end{figure}

While this configuration is effective and relatively simple to implement, the aperture mismatch between the secondary and tertiary objectives constitutes a significant drawback, limiting the performance and versatility of OPM. In fact, as the angle between the light sheet and the optical axis decreases, the fraction of fluorescence light coupled in the tertiary objective becomes smaller, reducing both signal intensity and imaging resolution \cite{kim2023_opm_review}. 
The problem is visually represented in figure \ref{fig. Apertures}, which shows the effective aperture angles for water immersion objectives on standard OPM for high and low numerical apertures. The scheme assumes the rarely implemented ideal scenario in which the tertiary objective has the same numerical aperture as the primary, which would require the presence of an immersion medium between the secondary and tertiary objective. Nonetheless, it can be observed that a significant portion of the aperture of the primary objective is not employed by the system, up to the extreme scenario of 0.5 NA objectives, in which the two apertures have no overlap, and no fluorescence photons can be detected.

As a result, only high numerical aperture primary objectives can be used effectively in standard OPM, and the total aperture and resolution of the system remain impaired from the re-projection setup.

This drawback can be mitigated by tilting the direction of propagation of fluorescence light through the use of a carefully positioned discontinuity in refractive index between the secondary and tertiary objectives. This can be achieved either through an immersion chamber with two separate liquids \cite{yang2019_immersion_remote_objectives} or through the use of a specialized axially asymmetric tertiary objective with null working distance\cite{sapoznik2020_snouty}. However, these modifications can be complex and expensive to achieve, and still require the use of high numerical aperture primary objectives in order to keep the angle between the light-sheet and the optical axis suitably large. OPM with low NA objectives has been proven possible through the use of a diffractive grating between the secondary and tertiary objectives \cite{hoffmann2019_diffractive_OPM}, which however introduces constraints on the maximum usable NA, and restricts use for multi-color imaging.

This manuscript presents remote focusing oblique plane microscopy (RF-OPM), an alternative approach to the oblique imaging problem. RF-OPM images the sample through a single objective of arbitrary numerical aperture, presenting no aperture mismatches in the detection path. Instead of using the standard OPM re-projection method, an electrically tunable lens (ETL)\cite{fahrbach2013_remote_focusing_lightsheet} is introduced in the back focal plane of the optical system, and a CMOS camera is positioned in the image plane. In this configuration, at any given time, only a narrow section of the oblique plane is in focus on the camera. The position of such narrow section can be dynamically shifted along the oblique plane by changing the focal power of the ETL. Through the linear modulation of the focal power of the lens in sync with the rolling shutter exposure of the camera, images of the oblique plane can be acquired entirely in focus. While limiting the rolling shutter exposure to a small sub-region of the detector significantly reduces the photon budget, the approach is already vastly employed in conventional light-sheet imaging, either to increase axial confinement in confocal light-sheet setups \cite{baumgart2012_confocal_lightsheet}, or to enable wide fields of view with high light-sheet numerical aperture \cite{dean2022_axially_swept_lightsheet}. These methods are generally considered a standard procedure when imaging very large optically cleared samples\cite{ueda2020_review_clearing}.

\section{Method}

\subsection{Working principle}
While RF-OPM can in principle be implemented using any form of incoherent remote focusing, including deformable lenses \cite{jiang2015_tunable_lens_remote_focusing}, deformable mirrors \cite{wright2021_deformable_mirror_remote_focusing}, Alvarez lenses \cite{bawart2018_alvarez_lenses} or optical elements mounted on fast accelerating translation stages \cite{botcherby2008_remote_focusing_actuators}, this manuscript reports an ETL based design. The ETL was chosen mostly for simplicity of implementation and reproducibility of the results, as it constitutes a relatively inexpensive and readily available device that has performance compatible with the experimental design.

\begin{figure}[ht!]
\centering\includegraphics[width=0.8\textwidth]{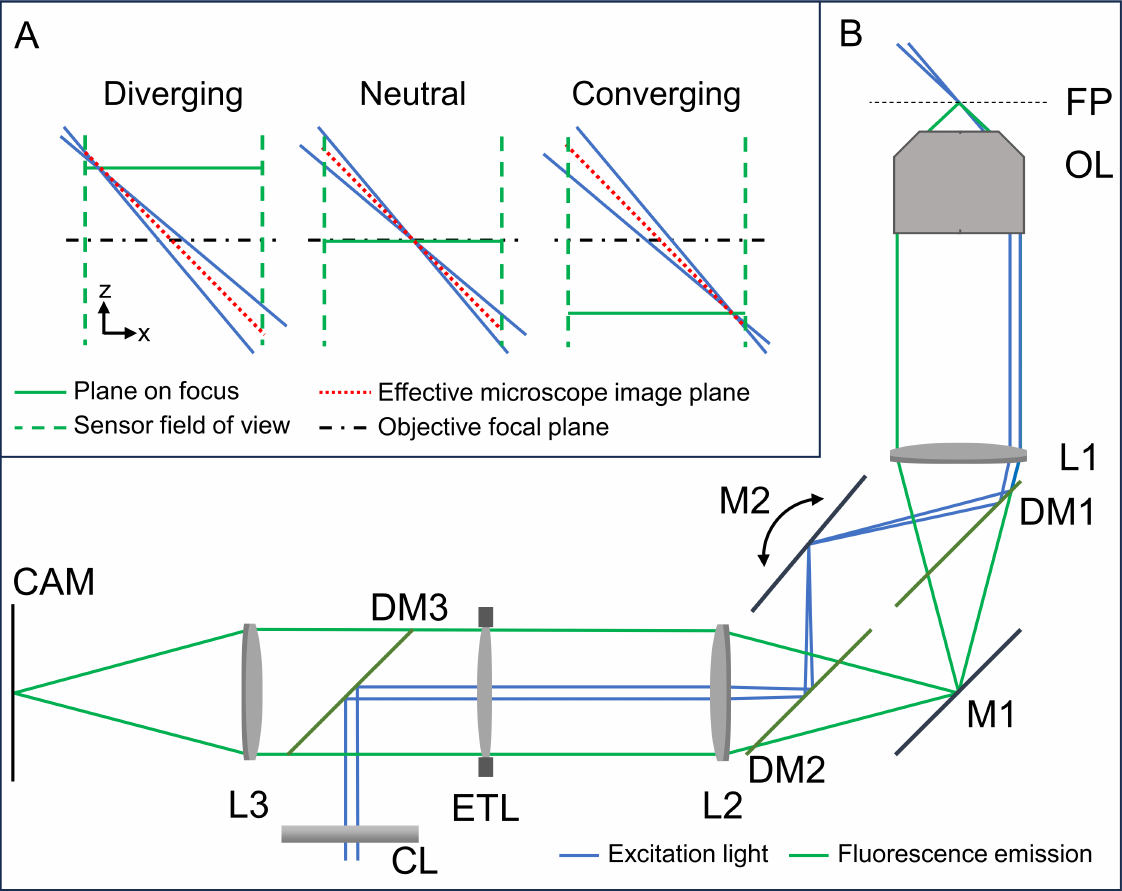}
\caption{\textbf{Method.} \textbf{A.} Scheme of the imaging geometry in the axial direction for different focal powers of the ETL.  \textbf{B.} Simplified schematic of the optical setup. FP - Objective focal plane, OL - Objective lens, PP - Pupil plane of the system, L1, L2 - 4f telescope, DM1, DM2, DM3 - Dichroic mirrors, M1, M2 - Image plane mirrors, ETL - Tunable lens, CL - Cylindrical lens, L3 - Tube lens, CAM - Camera detector plane.}\label{fig. Optical setup}
\end{figure}

A simplified representation of the RF-OPM optical setup is reported in figure \ref{fig. Optical setup}, panel B. The setup consists of a conventional epifluorescence microscope, with a 4-f telescope conjugating the objective's back focal plane with the aperture of an ETL. The conventional dichroic and filter set is then positioned between the ETL and the tube lens.

The laser light forming the light-sheet is focused by a cylindrical lens in the center of the ETL. Between the two lenses of the 4-f telescope, a pair of dichroic mirrors splits the excitation and detection paths, so that a manually adjustable mirror in the image plane of the system can be tilted to move the laser light at the edge of the objective aperture, achieving illumination along an oblique plane.

In this configuration, the light-sheet remains positioned along a fixed oblique plane, but its focus moves along the plane linearly with the focal power of the ETL, as shown in figure \ref{fig. Optical setup}, panel A. Neglecting the chromatic aberrations of the ETL, the camera always remains focused on the horizontal plane intersecting the light-sheet focus. As a result, at any given time, only a thin strip of the camera detects details in focus, with the lateral position of the strip shifting linearly with the ETL focal power. Using a rolling shutter sensor with a short exposure, it is possible to linearly modulate the focal power of the ETL so that the image is always in focus on the exposed pixels. The final output of the detector will therefore be an image of the light-sheet plane (indicated in red in panel A of figure \ref{fig. Optical setup}), fully in focus. Three-dimensional images can be acquired by either moving the sample through the light-sheet with a translation stage or, in principle, by conjugating the ETL plane with a galvanometric mirror, as is generally done in OPM.

It should be noted that the camera output obtained in this configuration consists of the projection of the oblique plane in the horizontal direction, and as such the magnification of the final image will be different for the $x$ and $y$ axes of the camera. An appropriate affine transformation should be applied to obtain an unwarped image.

\subsection{Field of view}

As discussed in the previous section, the magnification of the system is different along the axes parallel and perpendicular to the propagation direction of the light-sheet. When the light-sheet is correctly aligned at the edge of the pupil aperture, the angle of the beam from the optical axis is

\begin{equation}
\alpha = \arcsin{\left(\frac{{NA}_{obj}}{n}\right)}-0.5\arcsin{\left(\frac{{NA}_{ls}}{n}\right)}
\end{equation}

where ${NA}_{obj}$ and ${NA}_{ls}$ are the numerical apertures of the objective and of the light-sheet beam respectively, and $n$ is the refractive index of the sample.
For simplicity in the design of the system, this can also be approximated as

\begin{equation}
\alpha = \arcsin{\left(\frac{D_{obj}-R_{ls}}{2fn}\right)}
\end{equation}.

Where $f$ is the focal length of the objective lens, $D_{obj}$ is the diameter of the back aperture of the objective, and $R_{ls}$ is the radius of the light-sheet beam at the objective back focal plane.

Given the magnification $M$ of the system and assuming the use of a square detector of side $D$, the dimensions of the field of view of the acquired images is

\begin{equation}\label{eq:diagonal_fov}
{FOV}_{\bot}, {FOV}_{\parallel} = \frac{D}{M},\frac{D}{M \sin{\alpha}}
\end{equation}

where ${FOV}_{\bot}$ is the size in the direction perpendicular to the propagation of the light-sheet, and ${FOV}_{\parallel}$ is the one along the propagation direction of the light-sheet.
As a consequence, the total axial distance covered by the field of view is 

\begin{equation}\label{eq:axial_fov}
{FOV}_{axial} = \frac{D}{M \tan{\alpha}}
\end{equation}.

It is important to notice that the final pixel size along the propagation direction of the light-sheet increases with the tangent of the angle, approaching infinity as the light-sheet becomes vertical. In principle, an extremely wide field of view can be achieved in the axial direction by using a very low numerical aperture objective or by moving the light-sheet beam towards the center of the objective's aperture, but very low pixel sampling would be obtained in this situation.

A second limit to the achievable axial size of the field of view is given by the effective dioptric power range of the ETL, and its ability to linearly modulate it at a frequency compatible with the desired frame rate.
To ensure the most effective use of the ETL focal power range, the telescope conjugating it to the objective back focal plane should precisely match the two apertures. In this configuration, the required achievable focal length (both converging and diverging) of the ETL is

\begin{equation}
{f}_{tl} = \frac{2M_{tel}^2f_{obj}^2}{{FOV}_{axial}}\label{eq. lens focal power}
\end{equation}.

where $f_{obj}$ is the focal length of the objective lens and $M_{tel}$ is the magnification of the telescope between the objective and the ETL.

\subsection{Resolution and frame rate}
The estimation of the resolution of a standard OPM can be complex due to the dependence of the effective numerical aperture of the system on the angle between the secondary and tertiary objectives. A complete and exhaustive analysis can be found in Ref. \cite{kim2014_opm_resolution_sim}. In principle, RF-OPM images can achieve the nominal resolution of the objective in the oblique plane, while the resolution in the direction perpendicular to the oblique plane is given by the thickness of the light-sheet at its focus.

However, a limit to the performance is imposed by optical aberrations introduced by the ETL. When mounted horizontally, a conventional ETL introduces aberrations from the desired spherical phase with an RMS amplitude of less than $200 \, nm$, which can generally be neglected in light-sheet microscopy, since typical samples often introduce more severe aberrations \cite{furieri2023_muticonj_ao, wilding2016_lightsheet_ao}. However, when the ETL is operated at increasing frequencies, secondary modes other than pure defocus begin to be excited, which generally introduce spherical-like aberrations. For readily available commercial ETLs, such secondary modes exhibit a resonant frequency at approximately $400 \, Hz$, therefore precluding applications at extremely high frame rates. At more conventional light-sheet frame rates of tens of $Hz$, these effects can be considered negligible.

In addition to the excitation of secondary modes in the ETL, a second limitation to the maximum frame rate achievable in RF-OPM is given by the capability of the ETL to linearly modulate focal power over time. If a detector capable of alternating the direction of the rolling shutter at each frame is employed, the ETL can be synchronized with the detector through a triangular waveform. Triangular waveforms are relatively trivial to generate with ETLs, and nonlinear behavior is limited to short intervals between frames in which the direction of the scan is reversed. However, most detectors generally utilised in light-sheet microscopy are not capable of alternating rolling shutter direction at each frame.
In this case, the ETL must generate a sawtooth waveform, which presents a critical point at the discontinuity between frames. Since the ETL cannot instantly switch from a large positive focal power to a large negative one or vice versa, a significant interval between frames is necessary to allow the ETL to reset its position. Hence, as the microscope frame rate increases, the fraction of time in which the detector is actually exposed decreases, which could lead to an insufficient signal-to-noise ratio.

\subsection{Experimental setup}

Images were acquired on a custom RF-OPM setup. The setup is designed for a 60X, $1.1 \, NA$ water-dipping objective (LUMFL N 60XW, Olympus, Japan), with a back focal plane aperture diameter of $6.6 \, mm$. In order to prove the versatility of the method, additional experiments were performed with a 20X, $0.5 \, NA$ water-dipping objective (UMPlanFL N 20XW, Olympus, Japan).

The back focal plane of the objective is conjugated with the $16 \, mm$ aperture of the ETL (EL-16-40-TC-VIS-5D with ECC-1C controller, Optotune, Switzerland) through two lenses of $125 \, mm$ and $300 \, mm$ focal respectively (AC254-125-A and AC508-300-A, Thorlabs, USA), forming a 2.4X 4-f telescope. This resulted in perfect matching of the aperture of the 60X objective, while the numerical aperture of the 20X objective was cropped down by the ETL diameter to approximately $0.37$.

In order to image green fluorophores, long-pass dichroic mirrors with a cutoff wavelength at around $500 \, nm$ are used to split the path between the two lenses of the telescope. Dichroic mirrors with mm-scale thickness are employed (DMLP505L, Thorlabs, USA, $5 \, mm$ thick, and Di03-R488-t3-25x36, Semrock, USA, $3 \, mm$ thick) in order to minimize aberrations in the light-sheet path introduced by their curvature.

Excitation light is provided by a $40 \, mW$, $473 \, nm$ diode laser ($\lambda$-beam, RGB laser systems, Germany) coupled to a single-mode fiber. Light from the fiber is collimated to a beam diameter of $8 \, mm$ with a reflective collimator (RC08SMA-P01, Thorlabs, USA), focused through a $400 \, mm$ focal cylindrical lens (LJ1363RM-A, Thorlabs, USA) in a light-sheet conjugated to infinity by a $150 \, mm$ focal length lens (AC254-150-A, Thorlabs, USA), and then reflected through the center of the ETL by a dichroic mirror (DMLP505L, Thorlabs, USA). This configuration led to a final ratio of $0.18$ between the aperture diameters of the light-sheet and the objective at the back focal plane of the system.

An additional 0.5X 4-f telescope (AC508-200-A and AC508-100-A, Thorlabs, USA) is present in the fluorescence light optical path between the ETL and the tube-lens of the system. This addition serves two purposes: firstly, to extend the optical path length, allowing the horizontal mounting of the ETL; and secondly, to accommodate potential future upgrades, such as the integration of a galvanometric mirror at the system's back focal plane, which would enable faster dynamic imaging. A fluorescence filter (MF525-39, Thorlabs, USA) is present after the relayed back focal plane, and a $200 \, mm$ focal length tube lens (MXA20696, Nikon, Japan) is employed to conjugate the image plane of the system to an sCMOS camera (Orca Flash 4.0 v2, Hamamatsu, Japan), with a final effective magnification of 55.55X when using the 60X objective, and of 16.66X when using the 20X objective. Images were acquired on a 2048 by 1024 pixels subregion of the detector. The active region was cropped vertically to 1024 pixels, since the 60X objective showed significant spherical aberration outside this range, while the axial field of view achievable with the 20X objective was considered more than adequate for most available samples. Given the geometry of the setup, the final field of view of raw images for the 60X objective span $240 \,\mu m$ on the horizontal plane by $156 \,\mu m$ along the oblique plane, at an angle of $48^{\circ}$ with the optical axis, for a total axial range of $104 \, \mu m$. For the 20X objective, the field of view span $720 \, \mu m$ on the horizontal plane by $1020 \, \mu m$ along the oblique plane, at an angle of $20^{\circ}$ with the optical axis, for a total axial range of $958 \, \mu m$.

Due to the inability of the detector to image with an alternating direction rolling shutter, a sawtooth function is used for modulating the ETL focal power in time. The maximum usable imaging rate achievable is $50\, Hz$, with a duty cycle of exposure of approximately $60 \%$, while higher duty cycles can be achieved at lower frame rates, up to $90\%$ at $10 \, Hz$. Higher frequencies or better duty cycles could be achieved by vertically cropping the active region of the detector, reducing the axial size of the field of view.

Three-dimensional datasets were acquired by translating the sample horizontally with a servo-controlled actuator (M-405.CG, Physik Instrumente, Germany). A custom software script was used to correct the dataset shearing and stretching through an affine transform.

\section{Results}

\subsection{Microbeads imaging}
\begin{figure}[ht!]
\centering\includegraphics[width=0.8\textwidth]{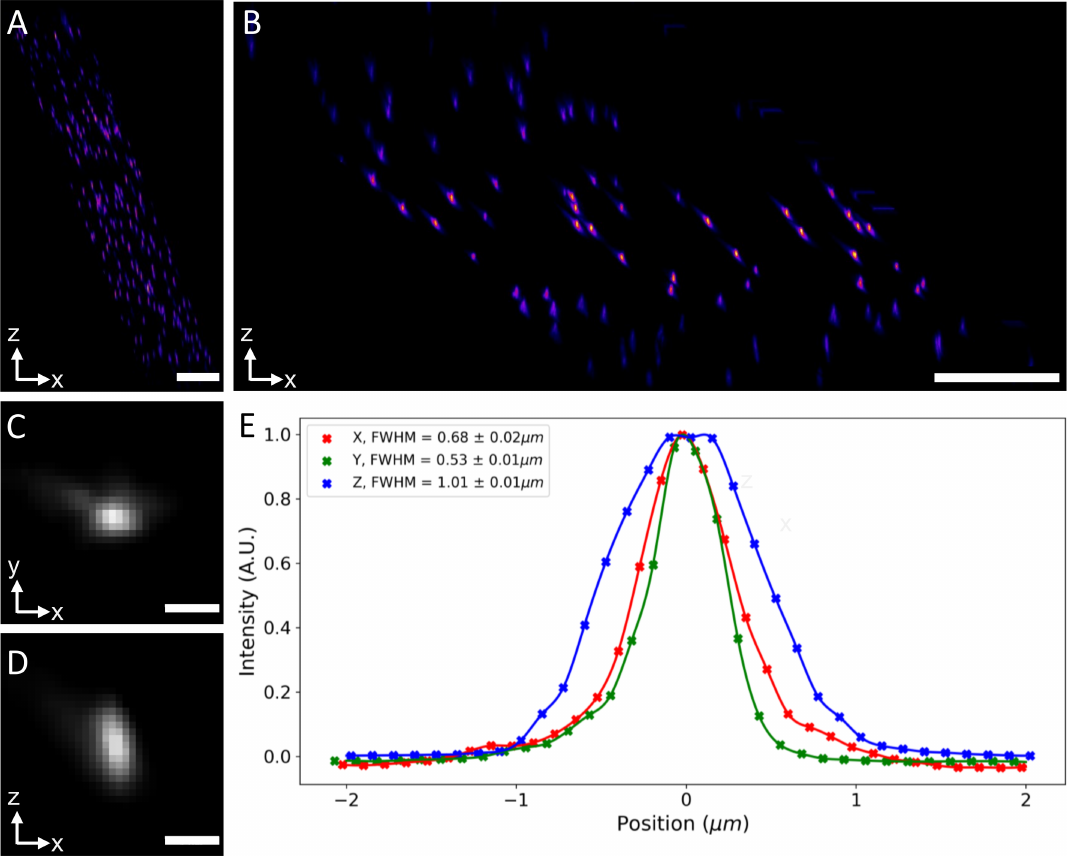}
\caption{\textbf{Microbeads imaging} \textbf{A} x-z maximum intensity projection (where z is the direction of the optical axis of the objective) over a $25 \, \mu m$ range in y of a microbead image, acquired with the 20X objective. Scale bar is $100 \, \mu m$. \textbf{B} x-z maximum intensity projection over a $100 \, \mu m$ range in y of a microbead image, acquired with the 60X objective. Scale bar is $25 \, \mu m$.  \textbf{C,D} images on a horizontal and vertical plane of a single microbead, acquired with a 60X objective. Scale bar is $1 \, \mu m$. \textbf{E} estimation of the full width at half maximum of the size of a single bead at the optimal working distance of the 60X objective. Lines are spline interpolation of data.}\label{fig. Beads}
\end{figure}

In order to evaluate the performance of the system, a $3 \, mm$ thick sample of $0.17 \mu m$ yellow-green fluorescent microbeads (P7220 PS-Speck Point Source Kit, Thermo Fisher Scientific, USA) embedded in agarose gel was imaged with the two objectives. Representative datasets are reported in figure \ref{fig. Beads}.

A first, important observation is that high aperture objectives are generally optimized for diffraction-limited imaging within a relatively short axial range of operation. This can be clearly observed in figure \ref{fig. Beads}, panels A and B, which show how the axial resolution of the system is conserved throughout the axial field of view with the 0.5 NA 20X objective lens, but is only optimal within a range of approximately $30 \, \mu m$ for the 1.1 NA 60X objective.
Within the optimal range of the objective, the full width at half maximum of the psf of the system is sub-micrometric laterally and around $1 \, \mu m$ axially. The resolving power is slightly worse than the nominal diffraction limit of the objective, and a slight star shape can be observed in the PSF. Both these effects are, most likely, due to slight astigmatism introduced in the detection path by the three dichroic mirrors utilised in the system. To avoid the introduction of vignetting in the system at high defocusing power of the ETL, large dichroics (2-inch diameter) were employed, which introduced non-negligible wavefront distortion. Moreover, the employed objectives are not optimized for use with a coverslip, which is necessary for imaging beads in an inverted microscope. A more optimized layout of the system using a coverslip-optimized objective and smaller, higher-quality dichroics should solve this issue.

\subsection{Mouse kidney imaging}
The main advantage of OPM systems over traditional light-sheet microscopy is its ability to image samples conventionally mounted on a microscopy slide.
In order to show this capability, a commercially available and fairly widespread sample, a $16 \, \mu m$ cryostat section of mouse kidney with immunofluorescence staining (Fluocells prepared slide \#3, Invitrogen, USA), was imaged with the setup.
\begin{figure}[ht!]
\centering\includegraphics[width=0.8\textwidth]{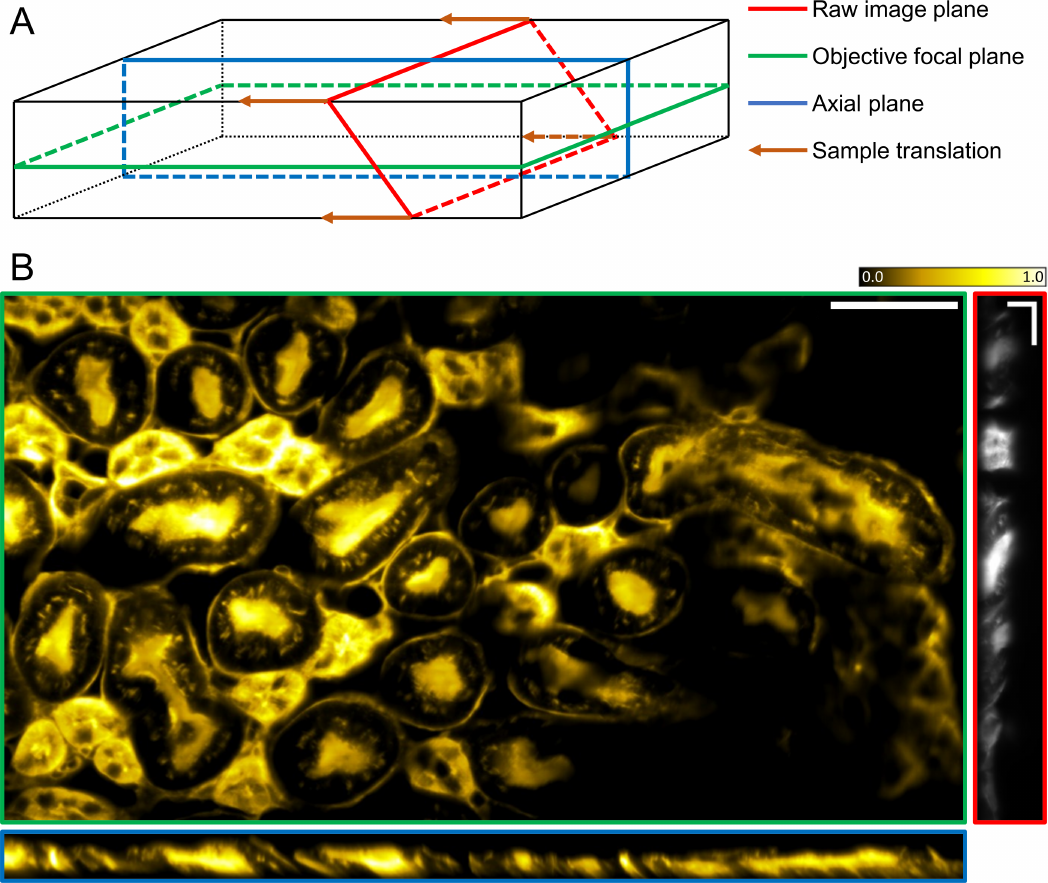}
\caption{\textbf{Prepared mouse kidney slide imaging.} \textbf{A.} Geometry of the acquisition, not to scale. The green plane represents the objective's focal plane, the red plane represents the raw field of view of the system, and the blue plane represents the axial field of view of the system. The orange arrow show the direction of the sample translation.  \textbf{B.} Representative images of the sample. Image border colors report their geometry referring to the scheme in panel A. Images in the horizontal and vertical plane are obtained through affine transform of the raw data. Scale bar for the horizontal and the vertical image is $50 \, \mu m$, scale bars in raw image are $20 \, \mu m$, and intensity scale bar is in arbitrary units.}\label{fig. Kidney}
\end{figure}
The glomeruli and convoluted tubules, stained with Alexa 488, were visible with the laser and filter set present in the setup.
Due to the thin nature of the sample, only the 60X objective was utilised.
Figure \ref{fig. Kidney} shows the acquisition geometry and a representative image of the sample. Raw planes were acquired at $50 \, Hz$, with a $0.4 \, \mu m$ spacing between planes, on a 2048 by 256 pixels sub-region of the detector.
The dataset shows good optical sectioning, and resolution performance comparable to those measured in the microbeads test. The images produced are also comparable to those reported in literature or on commercial microscopes documentation for the same sample. The main drawback of the current setup is the single-channel acquisition, which limits the amount of information that can be retrieved. Future upgrades with multi-edge dichroics could easily enable the use of multiple excitation wavelengths. The widely available dual rolling shutter feature of sCMOS cameras, together with the use of an image splitter, could also allow the simultaneous acquisition of two channels, doubling the acquisition pixel rate of the system. 

\subsection{Zebrafish imaging}
Tg(kdrl:eGFP)s843 zebrafish embryos, expressing a green fluorescent protein in the vascular structure \cite{jin2005_zebrafish_mutation} were imaged at 3 to 5 days post fertilization (dpf). The pigmentation was suppressed through 1-phenyl-2-thiourea (PTU) treatment \cite{karlsson2001_PTU}. To immobilize the larvae, 0.016\% tricaine anesthetic solution was used. The inverted nature of the microscope allowed convenient horizontal mounting of the larvae on a coverslip. A layer of 1\% w/v agarose was laid on the coverslip, in which a $0.8 \, mm$ wide groove cast with a 3D-printed comb was created to hold the larvae during imaging, as described in ref \cite{ahmed2020_zebrafishcomb}. Experiments were conducted with both objectives. High-resolution details of 3 dpf embryos were obtained with the 60X objective, while images of the full vascular system of a 5 dpf embryo were collected in a single sweep of the translation stage with the 20X objective. Images at 60X magnification were recorded at $20 \, Hz$ on the full 2048 by 1024 pixel active region, with a total data throughput of approximately $42\times10^6 \, pixels/s$, while 20X datasets were acquired at $50 \, Hz$ on a 2048 by 768 pixel active region, with a total data throughput of approximately $78\times10^6 \, pixels/s$. Three-dimensional datasets were captured by moving the translation stage, with horizontal spacing between frames of $0.4 \, \mu m$ for 60X images, and of $1 \, \mu m$ for 20X images.

\begin{figure}[ht!]
\centering\includegraphics[width=0.8\textwidth]{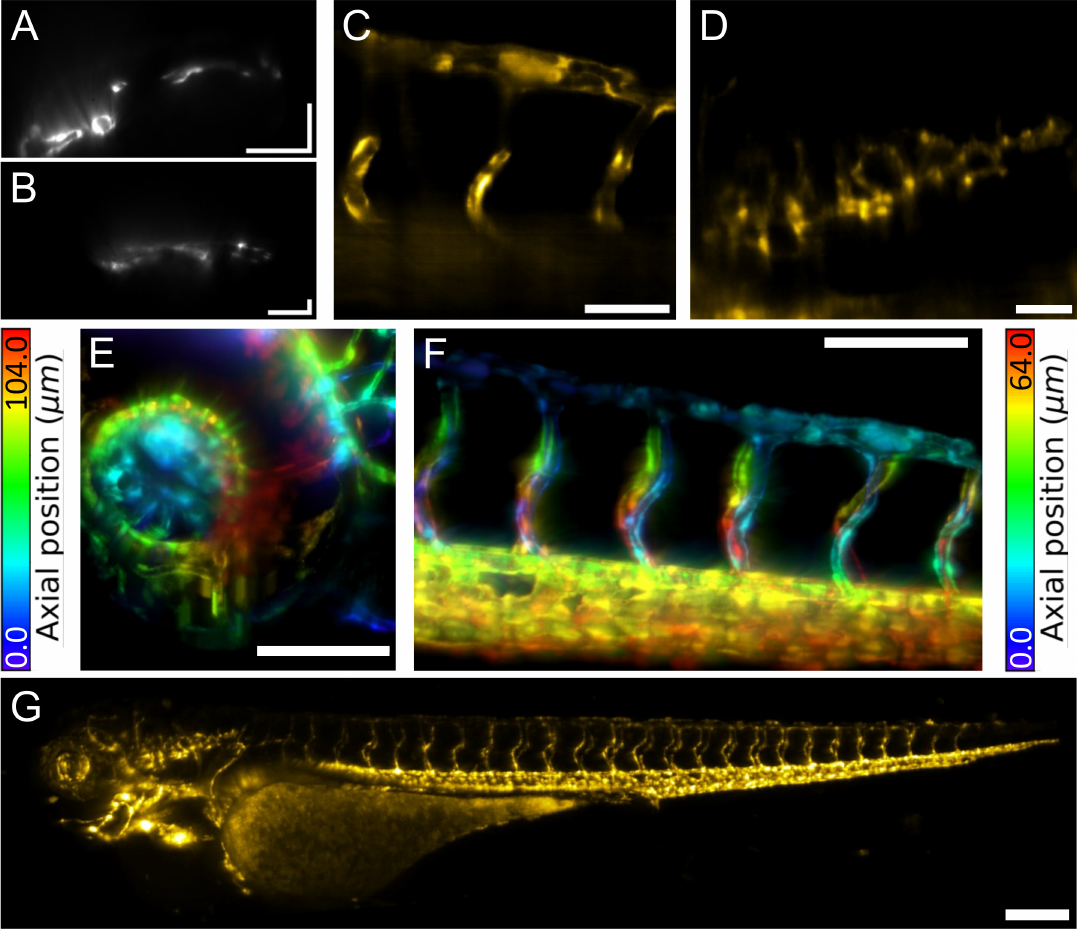}
\caption{\textbf{Zebrafish larvae imaging} \textbf{A, B.} Representative raw images from the detector during stack acquisition, with 60X and 20X objectives respectively. Scale bars are $50 \, \mu m$ in A and $100 \, \mu m$ in B. \textbf{C.} Typical two-dimensional lateral image after affine transform for a 60X image of the larva tail vasculature. Scale bar is $50 \, \mu m$. \textbf{D.} Typical two-dimensional lateral image after affine transformation for a 20X image of the larva vasculature. Scale bar is $50 \, \mu m$.\textbf{E, F.} Depth encoded projection of full dataset of the eye and tail vasculature respectively in 3 dpf larvae, acquired with the 60X objective at $20 \, Hz$, scale bars are $100 \, \mu m$. \textbf{G.} Maximum intensity projection of an entire 5 dpf larva acquired with the 20X objective at $50 \, Hz$, scale bar is $200 \, \mu m$.}\label{fig. Zebrafish}
\end{figure}

Typical datasets are reported in figure \ref{fig. Zebrafish}. Images acquired with the 60X 1.1 NA objective showed fine details in the vascular structure, both in the relatively clear tail sections and in the more complex and optically dense vasculature of the eye. The 20X objective produced, as expected, lower brightness and lower resolution images, but its wide field of view enabled the acquisition of images of the entire $3.5 \, mm$ long larva in a single sweep of the actuator in approximately one minute. Although the numerical aperture is lower, and therefore the quality of the images is not comparable, the field of view and pixel throughput of the microscope are comparable to state-of-the-art OPM with custom tertiary objectives \cite{yang2022daxi}.

\section{Discussion}
This manuscript introduces a novel approach to the visualization of tilted planes in OPM, employing an ETL and the detector rolling shutter to replace the reprojection setup of standard OPM. The proposed setup presents several advantages compared to a standard OPM, namely:
\begin{itemize}
    \item The system collects photons from the full aperture of the objective. While the gain in terms of effective signal is reduced by the need to have an exposure time shorter than the frame time to synchronize the rolling shutter with the ETL scan, this still enables imaging at higher resolution than standard OPM \cite{voleti2019_scape2}. Similar performance can be achieved with discontinuities in the refractive index between the secondary and tertiary objectives of standard OPM, but the presented method is arguably simpler and cost-effective.
    \item The proposed method allows great flexibility in the selection of the microscope objective. Unlike in standard OPM, low-aperture objectives can be used. Moreover, the objective can be rapidly switched during experiments, without the need to realign the optical system.
    The only limitation to the capability of operation with different objectives lies in the need to match the diameter of the back aperture of the objective with the diameter of the ETL. When the aperture diameter of the chosen objectives differs, as it did in the case of the ones employed here, the setup should be designed to either crop the pupil of the objectives with larger back apertures, or to underfill the ETL aperture when using objectives with smaller back apertures. Since, from equation \ref{eq. lens focal power}, reducing the magnification of the objective pupil at the ETL plane results in smaller axial shifts for the same lens current input, the presented setup was designed to always use the full aperture of the ETL and maximize the axial field of view when using the 60X objective, cropping the aperture of the 20X objective as a tradeoff. Different experimental needs may require different tradeoffs. Ideally, alternative remote focusing approaches with faster actuators and longer ranges may allow the use of the full aperture of any objective needed.
    \item The use of the same ETL for both excitation and fluorescence light ensures the light-sheet is always at its thinnest point in the region of the exposed rolling shutter. This greatly improves resolution at the axial edges of the field of view, and provides imaging of wider fields of view when compared to standard OPMs, or even traditional light-sheet microscopes with fixed cylindrical lenses. While this feature could be fully exploited in the presented data with the lower aperture objective, the narrow optimal axial range of higher performance objectives does hinder this capability of the system, due to the appearance of non-negligible spherical aberration when focusing further from the objective's focal plane. This position-dependent aberration could be reduced through the implementation of multi-conjugate aberration correction in the system \cite{furieri2023_muticonj_ao}.
\end{itemize}

\section{Funding}
The research has received funding from LASERLAB-EUROPE (grant agreement no. 871124, European Union's Horizon 2020 research and innovation programme) and the European Union's NextGenerationEU Programme with the I-PHOQS Infrastructure (IR0000016, ID D2B8D520, CUP B53C22001750006) “Integrated infrastructure initiative in Photonic and Quantum Sciences.”
The PhD student Alessia Brix was supported by the PhD program in Experimental Medicine of the University of Milan, Milan.

\section{Data availability} All the raw data for the images presented in the manuscript, together with software for affine transformation are available as an open dataset on Zenodo \cite{Zenodo_dataset}.

\bibliographystyle{unsrt}
\bibliography{sample}

\end{document}